# Lithography free method to synthesize the ultra-low reflection inverted-pyramid arrays for ultra-thin silicon solar cell

Anil Kumar[1], Divya Rani[1], Anjali Sain[2,3], Neeraj Joshi[1], Ravi Kumar Varma[1], Mrinal Dutta[4], Arup Samanta*[1,5]

[1]Department of Physics, Indian Institute of Technology Roorkee, Roorkee-247667, Uttarakhand, India
[2]Photovoltaic Metrology Section, Advanced Materials and Device Metrology Division, CSIR-National Physical Laboratory, New Delhi 110012, India
[3]Academy of Scientific and Innovative Research (AcSIR), New Delhi 110012, India
[4]India National Institute of Solar Energy, Gurgaon 122003, Haryana, India
[5]Centre of Nanotechnology, Indian Institute of Technology Roorkee, Roorkee-247667, Uttarakhand, India
*arup.samanta@ph.iitr.ac.in

**Abstract**
Silicon inverted pyramids arrays have been suggested as one of the most promising structure for high-efficient ultrathin solar cells due to their ability of superior light absorption and low enhancement of surface area. However, the existing techniques for such fabrication are either expensive or not able to create appropriate structure. Here, we present a lithography free method for the fabrication of inverted pyramid arrays by using a modified metal assisted chemical etching (MACE) method. The size and inter-inverted pyramids spacing can also be controlled through this method. We used an isotropic chemical etching technique for this process to control the angle of etching, which leads to ultra-low reflection, even < 0.5%, of this nanostructure. Using this specification, we have predicted the expected solar cell parameters, which exceeds the Lambertian limit. This report provides a new pathway to improve the efficiency of the ultrathin silicon solar cells at lower cost.

**Introduction:**

Low-cost silicon solar cell is the most important requirement in the photovoltaic industry. Major production cost of silicon solar cell lies in the use of thicker wafer to absorb the complete solar energy and the use of anti-reflection coating to reduce the surface reflection [1]. In general, multilayer antireflection coatings are used for the reduction of the surface reflection. However, this technique only reduces the reflection in a specific wavelength and at specific angles of incidence [2-9]. The solutions for the improvement of the efficiency and reduction of the cost of the silicon solar cells lies in the reduction of reflection of the solar cells [10-11].

Now-a-days, ultrathin silicon solar cell becomes an attractive research topic, since it has similar efficiency as the bulk crystalline silicon (c-Si) solar cell and it has significantly low production cost [12]. For this purpose, excellent light trapping is the most important point. Several important light trapping mechanisms in silicon have been theoretically proposed and experimentally developed also [13-28]. Various silicon nanostructures have been proposed for the light trapping mechanism, e.g., Nanowires [29-30], nanospheres [31], nanodome [32], nanopyramids [33] and inverted pyramids [34-37].

In general, surface silicon texturing with nanopyramids was performed by alkali anisotropic chemical etching (KOH or NaOH) [33]. However, the surface reflectivity is 10-12% for the pyramids textured silicon, which is close to the light trapping characteristic [38]. On the other hand, silicon inverted pyramid arrays are most promising structure for high-efficiency solar cells. [39] This structure has the tendency to reduce the reflectivity to extremely low level as well as the high absorption in the ultraviolet, visible and near infrared region [40-42]. On the other hand, surface area of this structure is far lower than the other silicon nano-structures like nanowire and nanoholes, which ultimately reduces the surface recombination, the main drawback of nanostructure silicon [28].

In general, silicon inverted pyramids (IPs) are synthesized by anisotropic metal assisted chemical etching method [34-37] or lithography assisted method [43-44]. Most of the cases, observed reflection is ~5% and improvement in the absorption in the UV and near-infrared region [45]. In this report, we have developed a lithography free method for the synthesis of the size and inter-pyramid-space controlled silicon nano/micro-inverted pyramid arrays, which shows reflection less than 0.5 % in the entire UV-VIS-NIR range. We have also shown that absorption is close to 99% in the UV-VIS-NIR due to the modified photonic bandgap in this structure and it exceeds the Lambertian limit too.

**Experimental results and Discussions:**

Figures 1(a) – 1(h) schematically represent the process flow for the fabrication of the lithography free inverted-pyramid arrays (IPAs) on silicon wafer. The process could be applied for both mono-crystalline and poly-crystalline silicon (mc-Si and pc-Si). Figure 1(a) shows the synthesis of silica nano/micro particles of controlled diameters by using Stöber and modified stöber method [46]. Figure 1(b) shows the formation of a monolayer on top of the silicon wafer by dip coating process. Figure 1(c) represents inter silica particle separation using reactive ion etching method. Figures 1(d) exemplify the metal (Au, Ag or Cu) coating on top side.

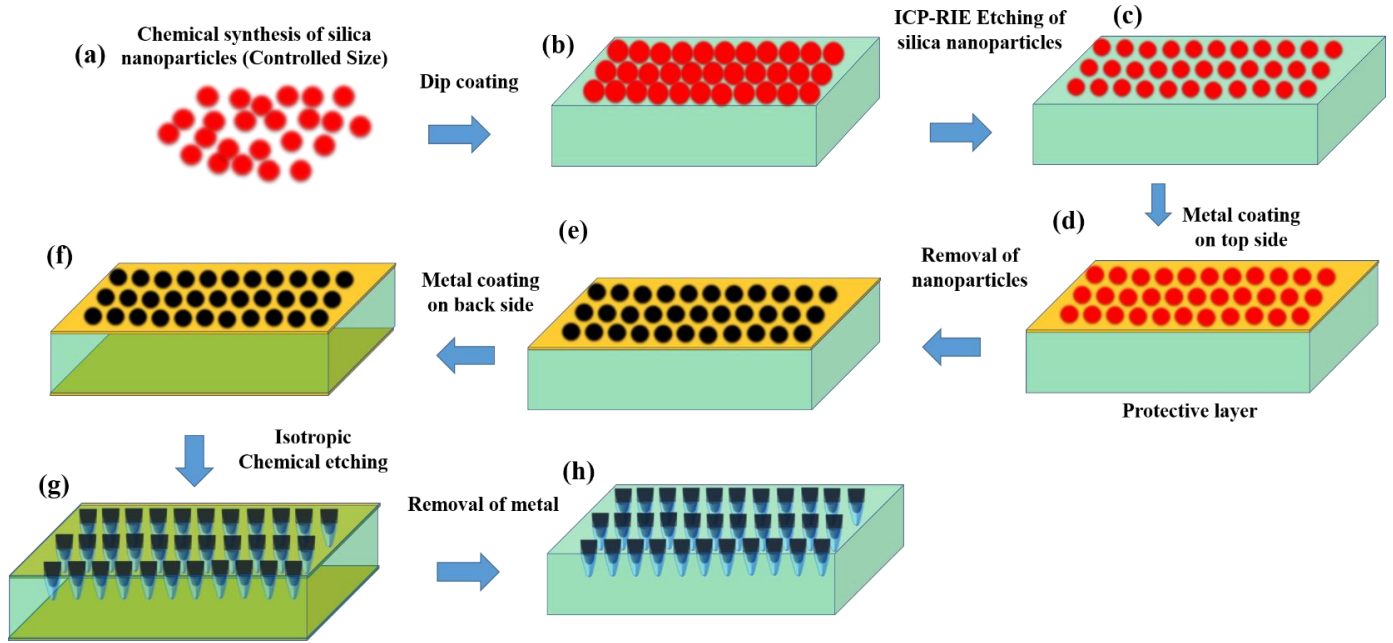

Figure 1: Schematic process flow chart for the fabrication of inverted pyramid arrays on mono-crystalline and poly-crystalline silicon wafer. We have developed process using isotropic chemical etching technique, which can be equally applicable for both mono-crystalline and poly-crystalline silicon.

Figure 1(e) shows the silicon wafer after the removal of silica particles by the ultra-sonication method. We have coated the metal in the back side of the wafer for the protection of silicon during the chemical etching process, as schematically presented in Fig. 1(f). Figure 1(g) shows the schematic surface after isotropic chemical etching using HNA. Figure 1(h) represents the final inverted pyramid arrays texture on silicon wafer after the removal of metal coating.

Firstly, we have synthesized silica nano/micro particles of various diameters using Stöber and modified Stöber method [46]. In this process, we take a desired concentration of ethanol and ammonium hydroxide in a polyethylene beaker and rotated the solution in magnetic stirrer. After 10 to 15 minutes, DI water would be added in the solution. After additional rotation of the solution for 10 - 15 minute, TEOS was added and leave this solution for 6 to 8 hours. Finally, the solution was kept as it is for 24 hours to precipitate the particles. Particles cleaned by multiple times centrifuge and sonication using ethanol as a solvent. We have synthesized various sizes of silica particles, as we can see in the figures 2(a - f). We have fabricated silica particles with the average size in-between 125 nm and 1100 nm as presented in figures 2(a - f).

Now for the preparation of monolayer on the top surface of silicon wafer, p-type monocrystalline Si (mc-Si) wafer should undergo standard RCA cleaning process and then to make it hydrophilic, we dipped it in the piranha solution for 24 hours. Thereafter, the monolayer of silica particles was prepared on the mc-Si wafer by dip coating method by managing various parameters like solution concentration, temperature, withdrawal speed and immersion time. In this process, 2.5 wt% concentration of silica particles in the solution of ethanol and DI water in the ratio of 3:1, immersion time of 4 minutes and withdrawal speed of 3.5 mm/sec were used. We can see from Figs. 3(a) and 3(b) that we have obtained monolayer of silica particles of 600 nm and 1100 nm, respectively.

For the control of the inter-inverted pyramid's gap, a proper inter-particle gap must be created. In this case, we have used appropriate plasma chemical etching by inductively-coupled Reactive Etching Process (ICP-RIE). We have created the plasma by using $CF_4$ gas with flow rate of 20 SCCM, RF power of 200 W, pressure of 2 Pa and substrate temperature at room temperature. The said plasma can only etch the $SiO_2$ with etching rate of 45 nm/min and with a negligible etching of silicon underneath. This etching process was used to make isolation of silica particles after making monolayer, as shown by the SEM images in Figs. 3(c) and 3(d) for 600 nm and 1100 nm particles, respectively. The gap created between the particles is ~ 200 nm, which can also be controlled depending of the requirement. Reactive ion etching creates uniform gap around particles that will help to creates uniform order in inverted pyramid arrays.

After that, we have deposited the gold (Au) by sputtering method on the top side. The thickness of the metal should be 1-10 nm. Metal coating provide resistance when isotropic etching will take place. After the metal coating, silica particles removed from the substrate by ultra-sonication method. In this process silicon wafer can be exposed at the position of the silica particles keeping other position metal coated, as shown in Fig. 4. Where black circles show non coated region and the whitish region is the metal coated part.

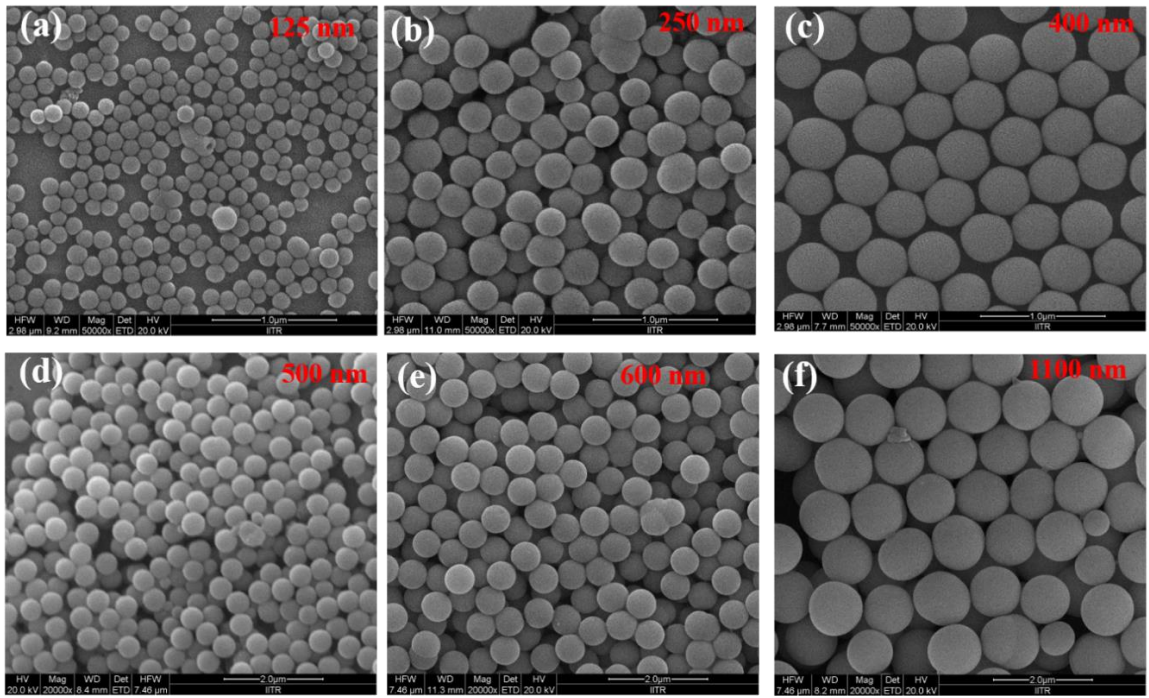

Fig.2: Scanning electron images of silica particles of different dimensions: (a) 125 nm, (b) 250 nm, (c) 400 nm, (d) 500 nm, (e) 600 nm and (f) 1100 are presented. These silica particles are prepared by optimizing the chemical composition following the Stöber and modified Stöber method. We have optimized the process to create the silica particle from 20 nm to 2000 nm.

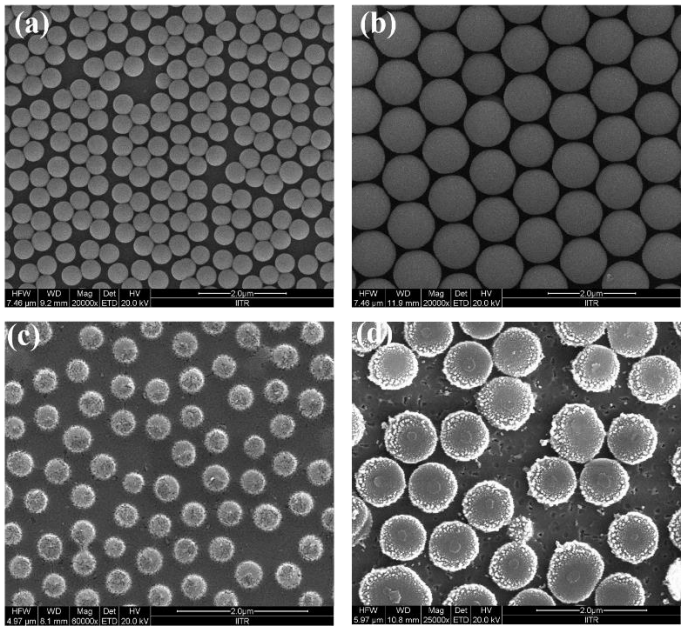

Figure 3: (a) and (b): SEM image of the monolayer formation of silica particles of different dimensions 600 nm and 1100 nm, respectively. (c) and (d): SEM image of the inter-particle separated monolayer created by ICP-RIE etching process for the above said particles.

Thereafter, we have deposited metal on the back side of the wafer for the protection of the wafer during the isotropic etching process. For the fabrication of inverted pyramids, these samples were immerged in optimized HNA (HF: $HNO_3$: Acetic acid) composite acid solution for isotropic etching of the exposed silicon area. The etching rate of the wafer using HNA is optimized in the range of 0.05 - 1.5 μm/min for low etching rate using composition ratio of HNA in 1:7:2, 2:6:2, 2:6:3 and 2:7:2. Au coated area remained as it is during this process. This isotropic etching process is suitable candidates for the etching of both mono-crystalline silicon and poly-crystalline silicon wafers such that only non-coated area is uniformly etched to make inverted pyramid arrays. After this process, Au can be removed by emerging it into acqua regia solution ($HNO_3$: HCl= 1:3).

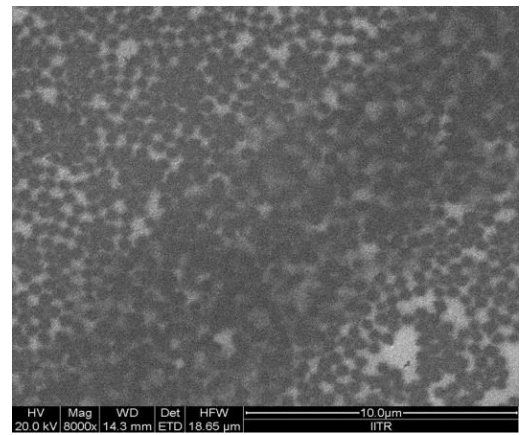

Figure 4: SEM image of the gold coated sample after he removal of silica particles.

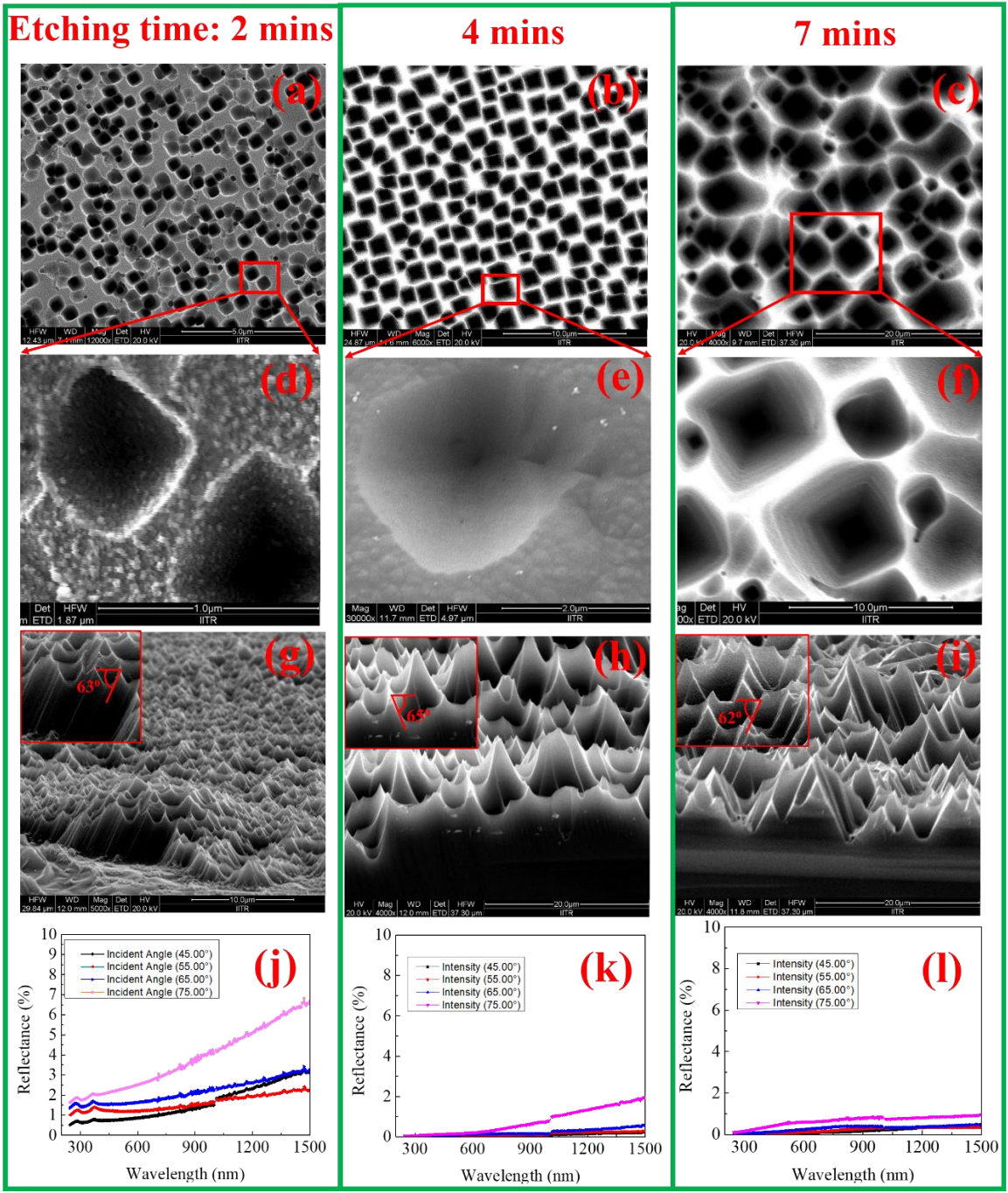

Figure 5: (**a-c**): SEM images for the IPs prepared from 1100 nm silica particles with etching time of 2 mins, 4 mins and 7 mins, respectively. (**d-f**) represented the zoom in IP structure of these materials. (**g-i**) represents the cross sectional view of the materials presented in Figs. (a-c). Insets: show the zoom in cross sectional images with the measured angle of etching. (**j-l**): Represent reflectance of samples measured with different angle of incidents starting from 45º to 75º.

Now the desired inverted pyramid arrays texture on the top of the silicon wafer remains left. Figures 5(a), (b) and (c) represent the inverted pyramid arrays structures prepared from 1100 nm silica-particle-coated silicon wafer following the process steps mention in Fig. 1 with etching time of 2 mins, 4 mins and 7 mins, respectively. Figure 5(a) is the representative data for less etched sample to keep gap between IPs, while Fig. 5(b) represents the perfectly/slightly-over etched sample as the size of IPs are as that of silica bids and Fig. 5(c) symbolizes over etched sample since the size of the IPs are larger than the size of the bids and a new epi-centers of inverted-pyramids have also been activated, which leads to uncontrolled size inverted pyramids. Figures 5(**d**) **to (f)** represent the zoom in IP structure of these materials. In this way, the present process can provide tunable size and inter-inverted-pyramids gap controlled textured silicon depending on the initial size of the silica bids, inter-particle gap and etching time.

Cross sectional view of these samples are presented in Figs. 5(g-i). The zoom in images of these structures are presented in the inset of their respective images. 2 mins etched sample has the etching angle of ~63⁰, while 4 mins etched sample has etching angle of ~65⁰ and 7 mins has the etching angle of ~62⁰, as presented in the insets of Figs. 5(g-i), respectively. The slight variation of the etching angle is due to slight change in the composition of HNA. In general, IPs are formed by anisotropic etching with the etching angle of 57.4⁰. Higher etching angle is obtained due to composition of HNA. This condition leads to low reflection in these samples due to additional reflections and absorptions of the incident light. The reflection measured by spectroscopic ellipsometry for different incident angles are presented in Figs. 5(j-l). A huge improvement in the reflection is observed in 4 mins and 7 mins etched samples. 4 mins etched sample shows reflection less than 1 % and 7 mins etched samples shows reflection is less than 0.5% in the entire range of 300 nm to 1500 nm.

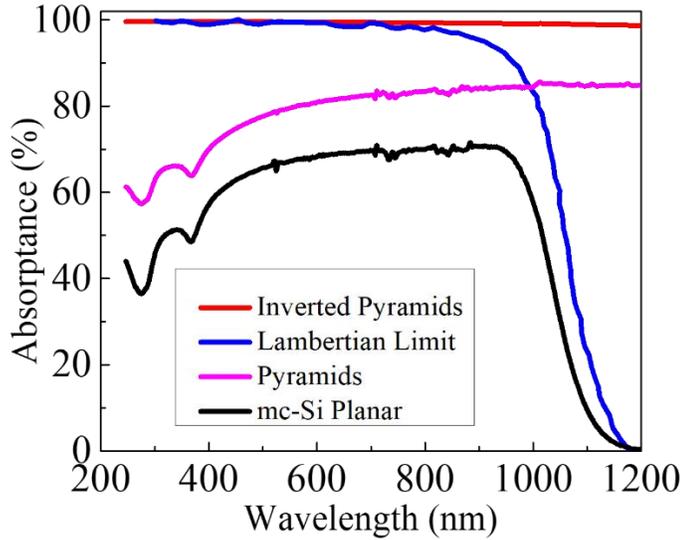

Figure 4: Absorptance spectra for Inverted Pyramid Arrays for the sample presented in Fig. 5(c), Pyramids Arrays prepared by standard alkali anisotropic etching method, planar mc-silicon wafer measured from Spectroscopic Ellipsometry at 75° incident angle and Lembertian limit Absorbtance for different samples were calculated by modeling the ellipsometric data.

We have also measured the absorption coefficient by modeling of ellipsometry data and from these data we have calculated the absorptance for these samples. One example for the absorbtance data for 7 min etched sample is presented in Fig. 6. For reference, we have plotted the absorptance for mc-Si wafer, pyramids textured wafer and Lembertian limit in the same graph. We can understand from Fig. 6 that our IPAs textured silicon wafer cross the Lembertian limit in the NIR range. High absorptance in the NIR range is due to the creation of photonic crystal band gap inside the silicon bandgap by following the interference effect of multiple rays [47]. This result is unique still now. Larger absorption leads to enhance the short-circuit current density, $J_{SC}$.

We have also calculated expected parameters of solar cell from these experimental data in the range of 300 nm-1100 nm and plotted the I-V curve in figure 7. We used solar cell parameters as presented below:
The short circuit current density is

$$J_{SC} = \int_{300nm}^{1100nm} \frac{e\lambda}{hc} I(\lambda) A(\lambda) d\lambda \qquad (1)$$

where, absorption coefficient, $A(\lambda) = 1 - R(\lambda) - T(\lambda)$
and R & T are reflection and transmission coefficients, other have their usual meanings.

$I(\lambda)$: Incident AM1.5 light intensity [48].

Reverse saturation current, $I_0$ is calculated using doping concentrations of $N_D$ and $N_A$ both are $10^{16}$ cm$^{-3}$, $10^{17}$ cm$^{-3}$, $3\times10^{18}$ cm$^{-3}$, and finally $N_D= 3\times10^{18}$ cm$^{-3}$ and $N_A=10^{16}$ cm$^{-3}$, which correspond to reverse saturation current, $I_0 = 5.566\times10^{-12}$ A, $6.66\times10^{-13}$A, $5.87\times10^{-14}$ A and $5.066\times10^{-12}$ A, respectively.

Open circuit voltage is given by:

$$V_{OC} = \frac{kT}{q} \ln(\frac{I_{SC}}{I_0} + 1) \qquad (2)$$

Where k is Boltzmann constant, T is temperature, q is electronic charge.

$$Fill\ factor, F.F = \frac{I_m V_m}{I_{SC} V_{SC}} \qquad (3)$$

wehre $V_m$= maximum voltage, $I_m$= maximum current.

$$Efficiency, \eta = \frac{F.F \times V_{OC} I_{SC}}{P_{in}} \qquad (5)$$

$P_{in}$= input power to the solar cell.

Figure 7 shows expected current-voltage characteristics of 7 mins etched IPAs textured silicon solar cell with different doping concentration. Here, we should mention that the $J_{SC}$ is 45.51 mA/cm$^2$, which is higher than the Lambertian limit. The effect of doping is presented for other parameters of solar cell in the graph. If we doped the both n-layer and p-layer with $3\times10^{18}$ cm$^{-3}$, the calculated open circuit voltage ($V_{OC}$) is 684 mV, Fill factor (F.F) is 84.54% and theoretical power conversion efficiency ($\eta$) is 26.35%. For this calculation, we didn't consider the surface passivation effect. Correction for surface passivation will further improve these parameters.

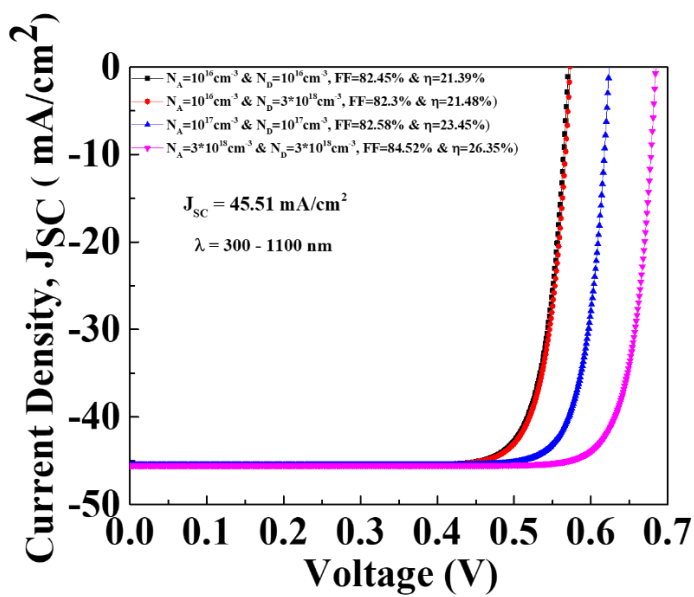

Figure 7: I-V curve for expected solar cells for 7 mins etched IPs along with all solar cell parameters.

The predicted efficiency of 26.35% is quite high for ultra-thin solar cell, which is comparable to the existing bulk crystalline silicon solar cell. This geometry has unique feature of multiple absorption and reflection, which minimize the reflection at surfaces and maximize absorption of light in the UV-VIS range while enhancement of absorption in the NIR range due to the photonic crystal's bandgap effect. Due to large absorption of incident light, large amount of electron hole pairs is generated, which is highly essential for the better performance of solar cell. In the present case, 10-15 µm inverted pyramid arrays can absorb full solar spectrum. This width is much less than the diffusion lengths of charge carriers which reduce carrier recombination losses. Hence, the present inverted pyramid arrays geometry will permit us to design efficient silicon solar cell at very low cost.

## Conclusion:

We have proposed a new method for the synthesis of size and gap controlled nano/micro inverted pyramid arrays on silicon wafer for fabrication of cost-effective ultra-thin silicon solar cells. The method includes the synthesis of silica nano/micro-particles with controlled diameter, coating of silica particles as a monolayer on the surface of mono-crystalline/poly-crystalline silicon substrate, controlled etching of particle by ICP-RIE plasma chemical process, metal coating onto the front and back sides of the silicon substrate, removal of masking particles, isotropic wet etching of the uncovered silicon substrate and finally cleaning of the residual metal or metal oxide from the surface of silicon wafer. The size of inverted pyramid arrays was controlled by etching rate and diameter of silica particles. In this IPAs textured silicon, we observed less than 0.5% reflection and more than 99% absorption in the UV-VIS-NIR range of the solar spectrum. Observation of ultra-low reflection in these structures is due to higher etching angel (>62º) of the IPs, which ultimately increases additional reflection and absorption of the incident light. The light trapping characteristics of these structure exceeds the Lambertian light trapping limit due to additional creation of optical resonance modes. Hence, the proposed inverted pyramid arrays of textured silicon open a new opportunity for the fabrication of low-cost silicon solar cell.

## Acknowledgement:

The authors acknowledge to Ministry of Education and CSIR, India for fellowship. The work is partially supported by DST-SERB (Project no: ECR/2017/001050), IIT Roorkee (Project no: FIG-100778-PHY) and DST-SERB Ramanujan Fellowship (Project no: SB/S2/RJN-077/2017) India.

## References:

1. L. C Andreani et al. (2019) *ADVANCES IN PHYSICS*: X, VOL. 4, NO. 1
2. Hu F, Sun Y et al. (2017) *Sol Energy. Mater. Sol. Cells* 159: 121–127
3. Eisenlohr J et al. (2016) *Sol Energy. Mater. Sol. Cells* 155:288–293
4. Lee Y, Kim H et al. (2015) *Mater. Sci. Semicond. Proc.* 40:391–396
5. Cao F et al. (2015) *Sol Energy. Mater. Sol. Cells* 141:132–138
6. Schindler F, Michl B et al. (2017) *Sol Energy. Mater. Sol. Cells* 171:180–186
7. Wu J, Yu P et al. (2015) *Nano Energy* 13:827–835
8. Yu P, et al. (2017) *Sci Rep.* 7:7696
9. Yu P, et al. (2017) *Sol Energy. Mater. Sol. Cells* 161:377–381
10. Vazsonyi E, et al. (1999) *Sol Energy. Mater. Sol. Cells* 57:179–188
11. Her T-H, et al. (1998) *Appl. Phys. Lett.* 73:1673–1675
12. A. Mavrokefalos et al. (2012) *Nano Lett*. 12, 2792-2796
13. M. D. Kelzenberg et al. (2010) *Nature Mater*. **9,** 239-44
14. Atwater et al. (2010) *Nat. Mater*. 9,205-13
15. Munday et al. (2011) *Nano Lett.* 11, 2195-201
16. Cao,L et al. (2011) *Nano Lett*. 10, 429-445
17. P.Campbell & M.Green (1987) *J. Appl. Phys*.62, 243-249
18. F. J. Haug et al. (2011) *J. Appl. Phys*. 109, 084516
19. A. Deinega et al. (2011) *JOSA A* 28, 770-777
20. P. Bermel et al. (2007) *Opt. Express* 15, 16986
21. Yoo JS, Parm IO et al. (2006) *Sol Energy Mater Sol Cells* 90:3085–3093
22. Zhong S, et al. (2015) *Adv Mater* 27:555–561
23. Yue Z et al. (2014) *Appl Phys A Mater Sci Process* 116:683–688
24. Oh J, et al. (2012) *Nat Nanotechnol* 7:743–748
25. Liu Y et al. (2012*) Small* 8:1392–1399
26. Savin H et al. (2015) *Nat Nanotechnol* 10:624–628
27. Lu Y-T, Barron AR (2014) *J Mater Chem* A 2:12043–12052
28. Wang Y et al. (2015) *Sci Rep* 5:10843
29. M.Dong Ko et al. (2015) *Sci. Rep*. 5, 11646
30. Y. Zhang & H Liu (2019) *Crystals* 9(2), 87
31. J.Grandidier et al. (2012) *Phy.Status Solidi A*
32. R. Zhang et al. (2015) *AIP Advances* 5, 12709
33. Y. Zhang et al. (2015) *5[th] IC on AEMT*
34. L. Yang et al. (2017) *Sol Energy Mater Sol Cells* 166, 121-26
35. Xu Haiyaun et al. (2018) *Nanotechnology* 29, 015403

36. C. Zhang et al. (2018) *Nanoscale Res. Lett*. 13 : 91
37. Y. Wang et al. (2015) *Sci. Rep*. 5, 10843
38. S. Zhong et al. (2016) *Adv. Funct.Mater*, 26, 4768–4777
39. Eyderman et al. (2015) *J. Appl. Phys*. 118, 023103
40. Zang et al. (2018) *Nanoscale Research Latters*-018-2502-9
41. B. M. Kayes et al. (2005) *J. Appl. Phys*. **97,** 114302
42. M. D. Kelzenberg et al. (2010) *Nature Mater*. **9,** 239
43. Savas, T.A (1996) *Sci Technol. B*, 14, 4167-70
44. Ran Xu et al. (2019) *Elsevier*, 165, 1-6
45. M. Xne et al. (2020) *Nano Energy* 70, 104466
46. Xue Lie et al. (2014) *Integrated Ferroelectrics*, 154, 142-146
47. S. Bhattacharya & S. John (2019) *Sci Rep*. 9, 12482
48. See http://rredc.nrel.gov/solar/spectra/am1.5/ Air Mass 1.5